# Plasmonic Su–Schrieffer–Heeger chains with strong coupling amplitudes

*Benedikt Schurr[1,*], Matthias Hensen[2,*], Luisa Brenneis[2], Philipp Kessler[2], Jin Qin[1], Victor Lisinetskii[2], Ronny Thomale[3], Tobias Brixner[2,†], and Bert Hecht[1,†]*

[1] NanoOptics & Biophotonics Group, Experimental Physics 5, University of Würzburg, Am Hubland, 97074 Würzburg, Germany

[2] Institut für Physikalische und Theoretische Chemie, Universität Würzburg, Am Hubland, 97074 Würzburg, Germany

[3] Institute for Theoretical Physics and Astrophysics, University of Würzburg, 97074 Würzburg, Germany

*Shared first authorship

[†]Corresponding authors:
bert.hecht@uni-wuerzburg.de (ORCID 0000-0002-4883-8676);
tobias.brixner@uni-wuerzburg.de (ORCID 0000-0002-6529-704X);

**Abstract**

Plasmonic many-particle systems with precisely tuned resonances and coupling strengths can exhibit emergent collective properties governed by universal principles. In one-dimensional chains with alternating couplings, known as Su–Schrieffer–Heeger (SSH) systems, this includes the formation of topologically protected mid-gap modes whose intensities localize at the chain's ends. This subwavelength localization at optical frequencies is crucial for achieving strong coupling of mid-gap modes to two-level systems under ambient conditions, extending topological protection to hybrid light–matter states. Here, we have fabricated SSH chains from plasmonic nanoslit resonators with strong inter-resonator coupling. The alternating distance between the nanoslit resonators is controlled with sub-nanometer precision, enabling accurate prediction and experimental observation of topologically protected mid-gap modes via photoemission electron microscopy (PEEM). Our results open the path towards experimental realizations of two-dimensional photonic metasurfaces exhibiting higher-order topological modes that can be strongly coupled to single emitters and quantum materials at ambient conditions.

**Main: Introduction: Plasmonic Su–Schrieffer–Heeger chains**

Topological photonics has emerged as a vibrant research field, offering new paradigms for controlling light at the nanoscale by leveraging concepts from condensed-matter physics[1]. Among these, the Su–Schrieffer–Heeger (SSH) model—a one-dimensional (1D) topological insulator—has been widely explored for its ability to host robust edge states protected by a topological invariant[2]. In terms of energy, these collective edge-state modes lie in a band gap of the system and are therefore also called mid-gap modes. Originally developed to describe electron behavior in polyacetylene[3], the SSH model has since been extended to photonic systems, where alternating couplings between resonators or waveguides produce mid-gap modes that although involving all particles of the chain exhibit spatially localized intensity at system edges[4–6]. Such mid-gap modes are particularly appealing for applications in robust light transport and quantum information processing owing to their resilience against perturbations. Plasmonic platforms, which enable deep subwavelength confinement of light, provide an ideal testbed for exploring topological phenomena at the nanoscale and offer the intriguing possibility to achieve strong coupling to quantum emitters at ambient conditions[7,8]. Furthermore, the mode localization of the mid-gap modes at the respective ends of the chain might be suited to establish a strong coupling of quantum systems by means of a topologically protected mode over distances larger than the working wavelength of the system[9].

Prior work was aimed at demonstrating plasmonic SSH chains and their unique mid-gap modes using linear disc chains[10]. There, however, overlapping particles and widely spaced particles were used to increase the coupling-strength contrast needed to energetically separate the mid-gap modes from the band-edge modes. This approach, however, compromises the identity of individual resonators so that no collective modes, and thus no mid-gap modes, can exist at all on the basis of identical chain links. Very recently, Yan et al.[11] have demonstrated near-field imaging of topological edge-state modes in plasmonic chains of gold nanodiscs connected by waveguides of varying widths. However, in this case, strong screening effects reduce coupling strength, leading to a narrowed bandgap and causing possible topological mid-gap modes to mix with energetically nearby trivial states. Notomi et al.[12] have used optical far-field microscopy to study plasmonic zig-zag chains exploiting alternating J- and H-like couplings between induced electric dipoles in neighboring metal discs. However, an increased light scattering signal at the ends of the chain, which can be associated with an edge-state mode, is only visible when, again, the gaps between the metal discs disappear and the individual resonators merge into a monolithic structure. At a distance of 17 nm between the metal discs and thus with an actual SSH chain of individual resonators, a scattering signal that is localized at the chain's ends could not be detected due to the low coupling strength and the associated spectral overlap of mid-gap mode and band edge modes. Furthermore, the optical diffraction limit did not allow a detailed characterization of mode structures. Near-field microscopy of short zig-zag chains was also reported[13]. But here, too, the coupling strength was limited due to a large metal disc distance of around 100 nm that does not allow to separate purported mid-gap modes and other modes. An indication of the outstanding feature of the nontrivial topological phase, i.e., an alternating light intensity from disc to disc that decreases exponentially into the chain bulk[14], could not be demonstrated. In addition, the fact that the operation of zig-zag chains requires polarized light forbids to extend the concept beyond 1D systems. Further studies of plasmonic SSH systems involved parallel linear waveguide systems in which the interwaveguide couplings alternate[15]. While such systems have proven to be interesting platforms to demonstrate SSH functionality at mid-IR frequencies, they lack deep subwavelength field confinement at optical frequencies to enable strong coupling with quantum emitters.

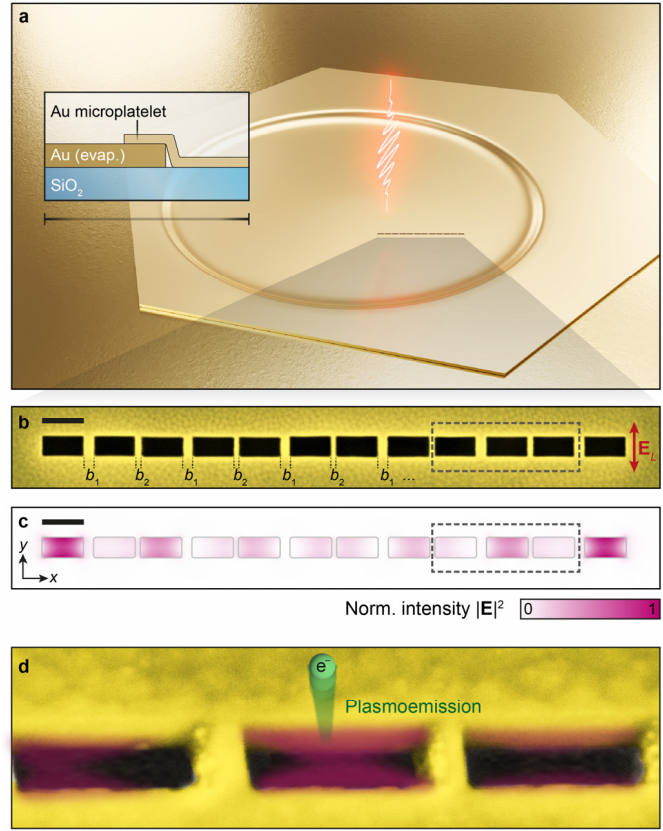

**Fig. 1: Plasmonic nanoslit SSH chain with nontrivial topology: Sample preparation, detection, and topologically protected mid-gap mode. a,** Chains of plasmonic nanoslit resonators written into a monocrystalline Au microplatelet using helium focused ion-beam milling. The microplatelet covers a hole in a vapor-deposited Au film created by a shadow mask. The nanoslit resonators therefore reside directly on a smooth glass substrate, while electric conductivity for PEEM is provided. Wide-field excitation (~270 µm spot diameter, much larger than a single SSH chain, flat wavefront) with ultrashort laser pulses leads to local plasmoemission of electrons according to the mode pattern that is imaged with PEEM. **b,** Top-view scanning electron microscopy (SEM) image (false color) of a nontrivial nanoslit SSH chain consisting of twelve coupled resonators separated by bridges with alternating widths, $b_1$ and $b_2$, as indicated. The polarization of the exciting electric field $\mathbf{E}_L$ (see artistic pulse in **a**) is oriented along the short axis of the nanoslits. **c,** Simulated near-field intensity of a mid-gap mode (COMSOL) exhibiting localized near-field intensity at the two outermost nanoslits. **d,** Side-view closeup SEM image of three nanoslit resonators, marked with a dashed rectangle in panels **b** and **c**, highlighting the quality of the structures. A purple overlay indicates the local near-field intensity distribution. An electron released via plasmoemission is indicated in green.

Here, we advance plasmonic topological Su–Schrieffer–Heeger (SSH) chains by mapping their subwavelength-localized, topologically protected mid-gap mode. Using photoemission electron microscopy (PEEM), we achieve nanoscale-resolved imaging of plasmonic modes, revealing how the modal structure depends on the system's geometry, i.e., its topological configuration (Fig. 1). The SSH chains consist of coupled Babinet-type nanoslit resonators, which allows the structure to be integrated and addressed in plasmonic circuits via plasmonic slot waveguides[16] or channel plasmon waveguides[17] with a minimum of material processing. Precise control over their spacing provides a large contrast between the alternating coupling strengths while always ensuring strong coupling to open up a sufficiently large energy gap. This allows us to experimentally confirm the presence of edge-localized mid-gap modes through their distinctive mode patterns that show excellent agreement with simulations. This study not only demonstrates a critical milestone in the control of subwavelength topological states but also lays the groundwork for future investigations of higher-order topological plasmonic systems[18], their coupling to quantum emitters[19], and their potential integration into quantum photonic technologies.

**Main: Sample design and preparation**

As fundamental building blocks of our plasmonic SSH chains, we use plasmonic nanoslit resonators[20,21], the Babinet counterpart of well-established nanorod dipole antennas[22]. This choice is particularly suited to meet the requirements of PEEM: At the light intensities required for photoelectron generation via multiphoton absorption, isolated nanostructures, such as nanorods, are prone to structural damage, and heat generated by absorption cannot be efficiently dissipated. Moreover, sufficient sample conductivity is essential to prevent charging effects which cause electromigration and discharges. By fabricating the nanoslits in a monocrystalline Au microplatelet directly connected to a vapor-deposited Au film (Fig. 1a), we achieve enhanced conductivity for both heat and charge carriers compared to isolated structures on conductive oxides such as indium tin oxide (ITO). Additionally, the use of nanoslits simplifies the manufacturing process, as only the material for the nanoslits needs to be removed and offers the possibility of integrating the SSH chain into plasmonic circuits via slot or channel plasmon waveguides. To achieve the required level of accuracy, we apply helium focused ion-beam milling (He-FIB, helium ion microscope, Zeiss Orion NanoFab) to monocrystalline Au microplatelets. The material's monocrystallinity ensures constant milling rates everywhere on the platelet. As a result, high fabrication accuracy and reproducibility can be obtained.

Fabricated SSH chains consist of twelve nanoslits (Fig. 1b) separated by alternating bridges of $b_1 = 24$ nm and $b_2 = 12$ nm width, with a standard deviation of below 1 nm (see Supporting Information, section S1). The nanoslits, resonant at the excitation laser wavelength of about $\lambda_L = 680$ nm (Fig.2b) were fabricated with a width of $w = 50$ nm and a length of $L = 100$ nm (Fig. 1b). The total depth of the nanoslits is determined by the thickness of the monocrystalline Au microplatelet and the incision depth in the glass substrate, which is caused by milling into the glass substrate after cutting through the gold microplatelet. The thickness of the monocrystalline Au microplatelet near its edge was measured with an atomic force microscope to be around 42 nm and is expected to be a few nanometers smaller at the nanoslit positions due to global material removal by the He-FIB process (more information in Supporting Information, section S1.1). To produce stable nanometer-sized bridges, we applied parallel He-FIB structuring of the entire SSH chain (see SI, section S1.2).

## Main: Strong coupling of nanoslit dimers

A prerequisite for the design of nanoslit SSH chains is the characterization of the coupling properties of end-to-end-aligned nanoslit resonator pairs depending on the bridge size $b$. To this end, we have carried out finite-difference time-domain (FDTD) simulations. The value of coupling strength $\hbar g$ (Fig. 2a) is determined by the spectral splitting $\Delta E = 2\hbar g$ of the two resulting nanoslit dimer eigenmodes in scattering cross-section simulations. To accurately determine the spectral position of the two eigenmodes, which can be difficult due to spectral overlap, we have performed two separate simulations in which the excitation conditions were chosen such that only one mode is selectively excited (Fig. 2b). As a source, one electric dipole was positioned in each nanoslit resonator oriented along the short axis of the respective resonator (Fig. 2a, inset). Note that due to the Babinet principle, the far-field excitation of a dipole-like resonance, i.e., the lowest-order cavity mode of a single nanoslit antenna, requires an electric field polarization along the short axis of the resonator[23]. The selection of the eigenmodes was then achieved by setting the relative phase between the dipoles to either 0 or π.

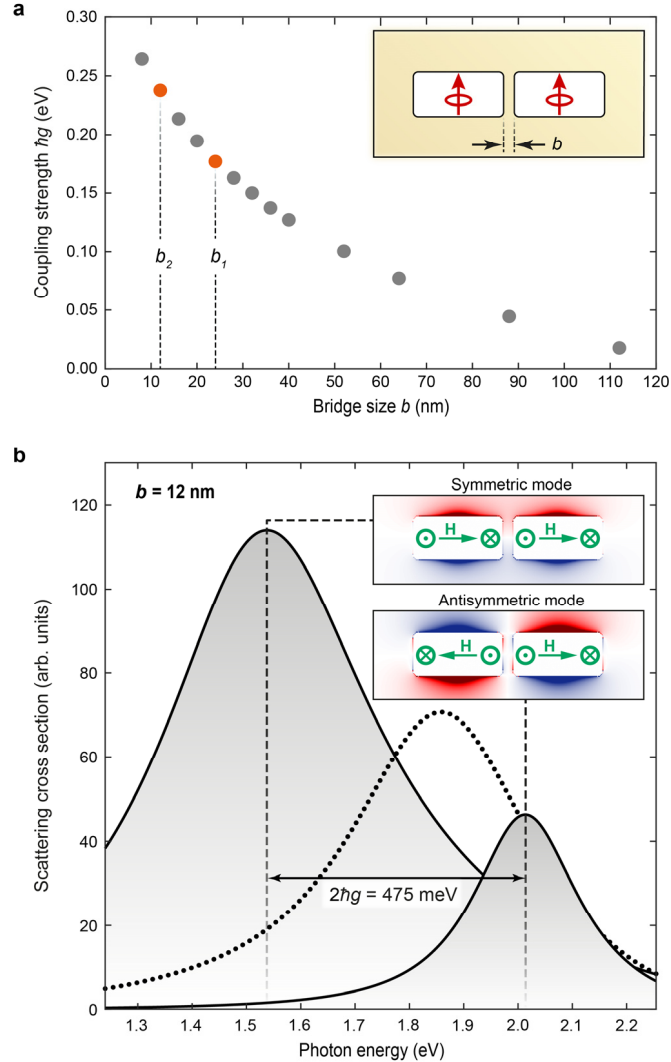

**Fig. 2: Coupling properties of a nanoslit dimer as retrieved from finite-difference time-domain (FDTD) simulations. a,** Coupling strength as a function of the distance $b$ between the nanoslits. As indicated in the inset, two electric dipoles, one positioned in the center of each nanoslit resonator, serve as the source to selectively excite the symmetric or antisymmetric hybridized eigenmode by setting the dipoles' relative phase to 0 or $\pi$, respectively. In both cases, a transmission monitor surrounding the structure measures the outgoing energy as a function of the oscillation frequency of the dipoles, resulting in a resonance curve. The value $\hbar g$ of the coupling strength is calculated via the spectral distance $\Delta E$ of the peak positions of the symmetric and antisymmetric hybridized eigenmode, i.e., $\frac{\Delta E}{2} = \hbar g$. The orange data points mark the coupling strength expected for nanoslit distances $b_1$ = 24 nm and $b_2$ = 12 nm of the fabricated structures. **b,** Resonance curves of the symmetric and antisymmetric eigenmode for a bridge size of $b$ = 12 nm (solid curves) as well as the monomer resonance curve (dotted line). The corresponding charge distributions of the dimeric eigenmodes (inset) were determined with built-in monitors and scripts of the commercial FDTD software. Green arrows mark the predominant direction of the magnetic field **H**. This field results for each monomer from the electric current flowing along the edges, driven by the electric field which can be represented by two magnetic dipoles, symbols $\odot$ and $\otimes$, which are aligned perpendicular to the figure plane and exhibit a phase difference of $\pi$.

In Fig. 2a, the coupling strength $\hbar g$ extracted from a series of such simulations is displayed as a function of the bridge size $b$. As expected for the coupling of plasmonic nanoresonators, the coupling strength increases nonlinearly with decreasing bridge size. For a bridge size $b < 10$ nm, the coupling strength reaches values of more than 250 meV. The data points marked in orange (Fig. 2a) indicate a coupling strength of 177 meV and 238 meV for bridge sizes $b_1 = 24$ nm and $b_2 = 12$ nm, respectively, which were used to fabricate plasmonic SSH chains. We note that these coupling strength values are close to or even higher than 10% of the single nanoslit resonance energy of about 1.859 eV (667 nm) and thus fall into the regime of ultra-strong coupling[24]. It can also be seen that for a bridge size $b > 100$ nm the coupling strength decreases by more than one order of magnitude compared to the value at the larger bridge size $b_1$ and thus we exclude an appreciable contribution from next-nearest-neighbor coupling in our SSH chain design in which the next-nearest nanoslit is located at a distance of $L + b_1 + b_2 = 136$ nm. This finding is significant since the original Su-Schrieffer-Heeger model[3] is based exclusively on nearest-neighbor coupling.

The simulated characteristic charge distributions of the two nanoslit dimer eigenmodes for a bridge of $b = b_2 = 12$ nm are shown as an inset in Figure 2b. Both eigenmodes exhibit charge accumulation along the short nanoslit axis. While charges of same sign accumulate on one side of the nanoslit dimer for the eigenmode at 1.537 eV (807 nm), charges of different sign accumulate on the same side of the nanoslit dimer for the eigenmode at 2.013 eV (616 nm). Consequently, only the red-shifted symmetric eigenmode can be efficiently addressed by far-field excitation with the polarization of the electric field along the short axis of the nanoslit resonators, whereas the blue-shifted antisymmetric mode can hardly be excited from the far field using plane waves at perpendicular angle of incidence. This behavior is in exact contrast to H-type coupling, i.e., the head-to-head configuration of electronic dipoles, in the context of molecular excitons. In excitonic H-type coupling, the energy of the bright transition is blue-shifted, just like in conventional plasmonic nanorod dimers, where the individual nanorods are aligned with their long axis parallel to each other. The apparent contradiction can be resolved by the fact that in the case of Babinet nanoresonators the coupling of magnetic dipoles plays a dominant role[25–27]. The electric currents that flow along the edges of a nanoslit monomer can be described by two out-of-plane magnetic dipoles that exhibit a π-phase shift, i.e., the overall current flow creates a magnetic quadrupole with an effective magnetic field **H** oriented along the nanoslit's long axis (Fig. 2b, inset). For the symmetric lower energy eigenmode, the magnetic dipole moments at the dimer bridge are aligned antiparallel to each other, whereas in the case of the antisymmetric mode, two magnetic dipoles aligned parallel to each other lead to a shift of the coupled mode to higher energies. In the context of plasmonic Babinet nanoresonators, the repulsion or attraction of magnetic poles takes over the role of the Coulomb interaction of electric charges in conventional plasmonic nanoresonators. In this sense, there is a J-type coupling behavior for the magnetic fields **H** pointing along the long nanoslit axis (Fig. 2b, inset). For the remainder of the manuscript, however, we stick to the description and nomenclature for the electric fields.

**Main: From nanoslit dimers to chains with topological features**

To show that the plasmonic SSH chain (Fig. 1b) with an H-type electric dipole configuration is in the topologically nontrivial phase, we will prove the existence of the associated mid-gap modes by a simulation-based mode decomposition. For this purpose, we will compare the plasmonic eigenmodes with the quantum mechanical eigenstates of a typical SSH Hamiltonian. We start with the quantum model and represent our system of twelve nanoslit resonators via a one-dimensional, spinless Hubbard Hamiltonian $\hat{H}$ of twelve equivalent two-level systems:

$$\hat{H} = \sum_{i=1}^{12} E_0 \hat{\sigma}_i^+ \hat{\sigma}_i + \sum_{\substack{i=1 \\ i\text{ odd}}}^{11} v(\hat{\sigma}_i^+ \hat{\sigma}_{i+1} + \hat{\sigma}_i\ \hat{\sigma}_{i+1}^+) + \sum_{\substack{i=2 \\ i\text{ even}}}^{10} w(\hat{\sigma}_i^+ \hat{\sigma}_{i+1} + \hat{\sigma}_i\ \hat{\sigma}_{i+1}^+), \qquad (1)$$

where $E_0$ is the monomer resonance energy, and $v$ and $w$ represent the values for the alternating coupling strengths between different monomer sites in units of energy. For each of the degenerate resonators within the chain, $\hat{\sigma}_i^+$ and $\hat{\sigma}_i$ are the associated fermionic creation and annihilation operators, respectively, considering only single excitations. In this limit, the diagonalization of the Hamiltonian (1), which consists of twelve equivalent monomers, results in a common ground state and twelve new eigenstates in the manifold of single excitations. This Hamiltonian corresponds to that of the original work[3] of Su, Schrieffer, and Heeger, except for the missing spin degree of freedom, the missing interaction details of hydrogen and carbon atoms in polyacetylene, and the first term, which serves as a gauge for the energy scale.

The systematics of the energy distribution of eigenstates in the manifold of single excitations crucially depends on the ratio of the coupling constants $v$ and $w$. For uniform coupling $v = w$, a homogeneous distribution of energy eigenvalues around the monomer energy $E_0$ is obtained, which would be gapless in the limit of infinite monomer sites. Note that only few states in the excited-state manifold can be excited efficiently from the ground state due to the oscillator strength redistribution[28]. In the case of excitonic H-type aggregates, these are the states of highest energy. Breaking the translational symmetry of the coupling constant via staggering, i.e., $v \neq w$, results in the splitting of the energy spectrum into two distinct bands separated by a gap. In addition, two topologically different phases can be distinguished[2]: A phase of trivial topology (TT) for $|v| > |w|$ and of nontrivial topology (NTT) for $|v| < |w|$.

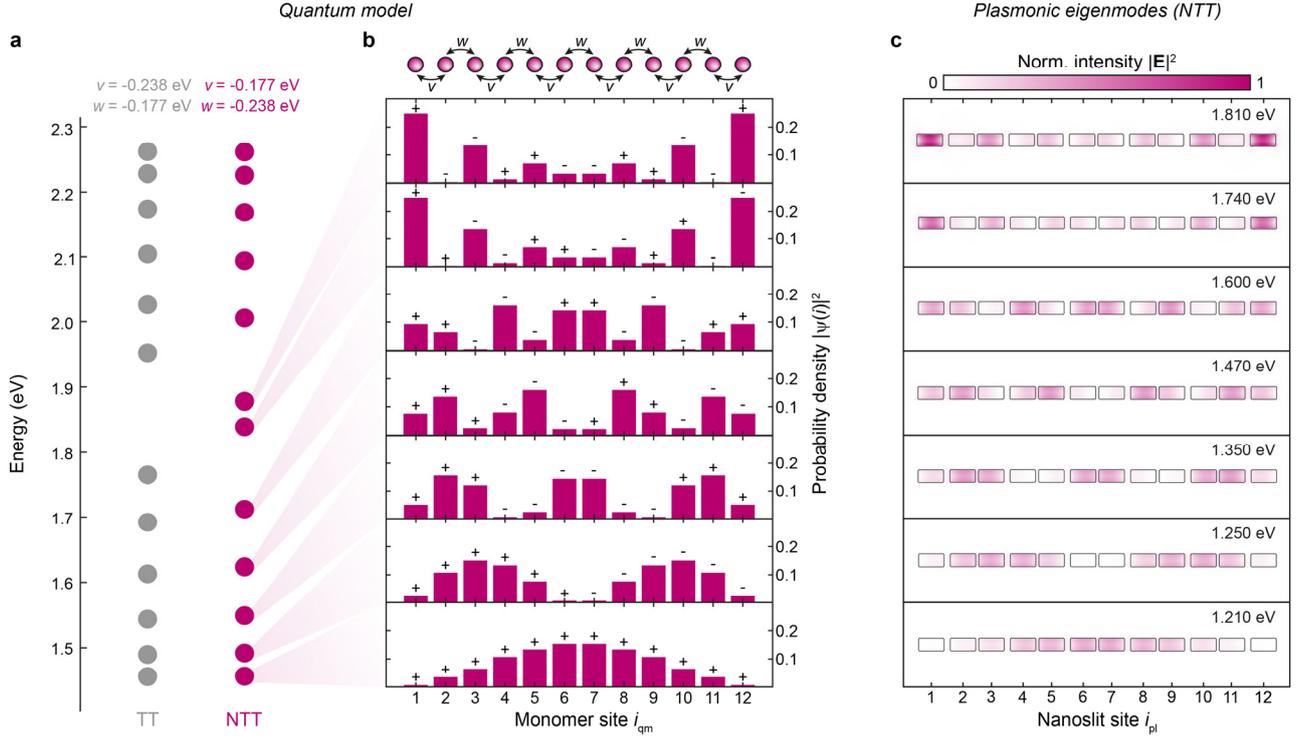

**Fig. 3: Modelling the nontrivial topological phase of the plasmonic nanoslit chain.** Comparison of the mode patterns of a quantum mechanical SSH chain (see equation (1)) consisting of equivalent two-level systems in the nontrivial topological phase to the corresponding mode patterns obtained via a quasi-normal mode analysis. **a,** Eigenenergies of the single-excitation manifold of twelve coupled resonators with equivalent excitation energies of $E_0 = 1.851$ eV. TT: trivial topology ($v = -238$ meV and $w = -177$ meV) and NTT: nontrivial topology ($v = -177$ meV and $w = -238$ meV) with coupling amplitudes $v$ and $w$. **b,** Probability density of the lower band and edge-state eigenfunctions of the quantum mechanical SSH chain in NTT configuration. The representation in the monomer site basis illustrates the localization of the excitations. **c,** Near-field intensity distribution of the plasmonic nanoslit chain, at half of its height, in NTT configuration ($b_1 = 24$ nm, i.e., 177 meV coupling strength, and $b_2 = 12$ nm, i.e., 238 meV coupling strength) for the seven lowest eigenmodes obtained by quasi-normal mode analysis (COMSOL).

Figure 3a shows the energy spectrum for the TT phase with $v = -238$ meV and $w = -177$ meV as well as for the NTT phase with $v = -177$ meV and $w = -238$ meV. Note that the values for $v$ and $w$ correspond to the coupling strength of a plasmonic nanoslit dimer for bridge sizes of $b = 12$ nm and $b = 24$ nm (see Fig. 2a). The negative sign accounts for H-type coupling of electric dipoles in the case of plasmonic systems. Moreover, the monomer energy is set to $E_0 = 1.859$ eV, i.e., the resonance energy of a single plasmonic nanoslit resonator (Fig. 2b). As compared to the TT phase, in the NTT phase the gap opens and two states appear in the middle of the band gap. These are the edge states, also called mid-gap states, which are topologically protected against perturbations, e.g., structural deformation.

In Figure 3b, we show the probability density $|\psi(i_{qm})|^2$ of having a single excitation in the NTT phase at the position of the $i$-th monomer. More precisely, $\psi(i_{qm}) = \sum_i |i\rangle\langle i|\psi\rangle$ is the wavefunction of a SSH chain eigenstate $|\psi\rangle$ expressed in the site basis $\sum_i |i\rangle\langle i|$, where $|i\rangle = |\ldots, g_{i-1}, e_i, g_{i+1}, \ldots\rangle$ describes the excitation of the $i$-th monomer. While the eigenstates in the lower (and upper) band are delocalized over the entire chain, the mid-gap states are strongly localized at the outer two monomers, $i_{\text{qm}} = 1$ and $i_{\text{qm}} = 12$ (see Supporting Information section S3 for all wavefunctions of the NTT and TT configuration). The localization can be increased by increasing the ratio $\frac{w}{v}$ since the probability density of these states decreases exponentially into the chain with a characteristic decay length of $\xi = \frac{1}{\log(w/v)}$, which is measured in number of sites[2]. It is interesting to note that although the mid-gap states are completely identical in terms of the probability density $|\psi(i_{\text{qm}})|^2$, they exhibit opposite parity: The sign of the real part of the monomer-resolved probability amplitude $\psi(i_{\text{qm}})$ is denoted by + and - signs in Fig. 3b (see Supporting Information section S3). As a result, the higher-energy mid-gap state has an even parity while the lower-energy mid-gap state has an odd parity. Consequently, only the higher-energy mid-gap state can be excited from the far field using a plane wave at perpendicular incidence. Due to the relatively short length of the chain, the two mid-gap states are the symmetric and antisymmetric superpositions of the real edge states. In the case of a very long chain, these real edge states would be localized either on the left or right side of the chain and would be degenerate in energy. Hence, the different parity of the mid-gap states is a consequence of the symmetric and antisymmetric superpositions of the left- and right-localized edge states. We also point out that the parity of the mid-gap states swaps when an additional dimer is introduced into the rather short chain. For each additional dimer, a new eigenstate is then added in the lower band so that the order of parity for the mid-gap states changes accordingly.

To verify that exclusive nearest-neighbor coupling is a valid assumption we now compare the probability density $|\psi(i_{\text{qm}})|^2$ of the SSH chain in the NTT phase with the quasi-normal modes of the fabricated nanoslit resonator chain determined using COMSOL (Fig. 1b, see Methods for details). According to the distance-dependent coupling strength (Fig. 2a), the staggered nanoslit bridges $b_1$ and $b_2$ should result in a nontrivial topological phase of the plasmonic system since $|v(b_1)| < |w(b_2)|$. As expected from the hybridization of twelve individual resonator modes, we find twelve eigenmodes for the nanoslit chain, of which we show the spatially resolved near-field intensity for the seven lowest energy modes in Fig. 3c (see Supporting Information section S4 for all modes). The agreement of the near-field intensity distribution obtained by solving Maxwells equations and the probability density of

the quantum model assuming the nearest-neighbor hopping of single excitations (Fig. 3b) is striking: Even if the plasmonic system does not consist of point-like chain elements but extended spatial structures with the corresponding spatially extended modes, the progression of the field intensity from nanoslit to nanoslit corresponds exactly to the progression of $|\psi(i_{\text{qm}})|^2$ confirming the validity of the pure nearest-neighbor coupling assumption.

Note that there is a slight difference between the eigenenergies of the plasmonic modes compared to the energies in the single-exciton manifold. Such a deviation is expected since the spectral splitting due to the hybridization of plasmonic modes is not symmetric with respect to the resonance energy of the individual nanoslit and is not considered in equation (1). Asymmetric splitting occurs due to the nonlinear Coulomb interaction between the electron densities involved[22]. For the present plasmonic system, the eigenenergies of the lower and upper band-edge states, as obtained from the mode decomposition, amount to $E = 1.600$ eV and $E = 1.920$ eV, respectively, resulting in a band gap of about $E_{\text{gap}} = 320$ meV (see Supporting Information section S4 for the spectral data of all modes). Expressed in relative scales, the size of the band gap is larger than 15% of the excitation energy of the nanoslit resonator.

Mid-gap modes in the fabricated nanoslit chains are expected to be found at eigenenergies of 1.740 eV and 1.810 eV, red-shifted with respect to the resonance energy of the individual nanoslit resonator (Fig. 3c). Note that a red shift of both mid-gap modes with respect to the resonance frequency of the individual resonator was also observed in a theoretical investigation of nanoparticle SSH chains[14]. The alternation from high to low electric field intensity from nanoslit to nanoslit, which is archetypal for these states[14], can be clearly recognized together with the exponential overall decrease of the intensity into the bulk of the chain. We therefore conclude that the fabricated plasmonic system is expected to be in the NTT configuration. Furthermore, we point out that, according to the correspondence of $|\psi(i_{\text{qm}})|^2$ and $|E(\mathbf{r})|^2$, the localization of electromagnetic energy can be increased at the outer nanoslits by increasing the ratio $\frac{w(b_2)}{v(b_1)}$, i.e., increasing the bridge size of $b_1$ compared to $b_2$. However, we intendedly avoid this regime since the extreme case of a large ratio $\frac{b_1}{b_2}$ simply results in a system of isolated nanoslit dimers and single nanoslit resonators with an unperturbed resonance energy at each end[10].

**Main: PEEM imaging of edge-state modes**

To experimentally prove the existence of topologically protected plasmonic edge-state modes in He-FIB-fabricated SSH chains, we combine aberration-corrected photoemission electron microscopy (PEEM), featuring a spatial resolution of about 3 nm[29], with a broadband NOPA system (full width at tenth maximum $\Delta\lambda$ = 120 nm or $\Delta E$ = 200 meV at $\lambda_L$ = 675 nm or $E_L$ = 1.837 eV, respectively) as excitation source to measure the mode structure shown in Fig. 3c. In PEEM, the dependence of the instantaneous photoelectron yield $Y(\mathbf{r}, t)$ on the $N$-th power of local intensity $|\mathbf{E}_{\text{loc}}(\mathbf{r}, t)|^{2N}$ of the electric near-field of the plasmons is used as imaging contrast that enables deep-subwavelength spatial and femtosecond temporal resolution[30]. Note that $N$ is the number of absorbed photons so that the imaging contrast for plasmonic excitations in the optical spectral range is based on nonlinear photoemission. Davis and co-workers had shown that it is even possible to reconstruct the full vectorial electric field for surface plasmon-polaritons with PEEM[31]. However, since we are only interested in the energy density distribution of plasmonic modes within the SSH chain, the intensity-dependent scalar value for the time-integrated photoelectron yield $Y(\mathbf{r}) \propto \int_{-\infty}^{\infty} |\mathbf{E}_{\text{loc}}(\mathbf{r}, t)|^{2N} dt$, i.e., a static PEEM image, is sufficient.

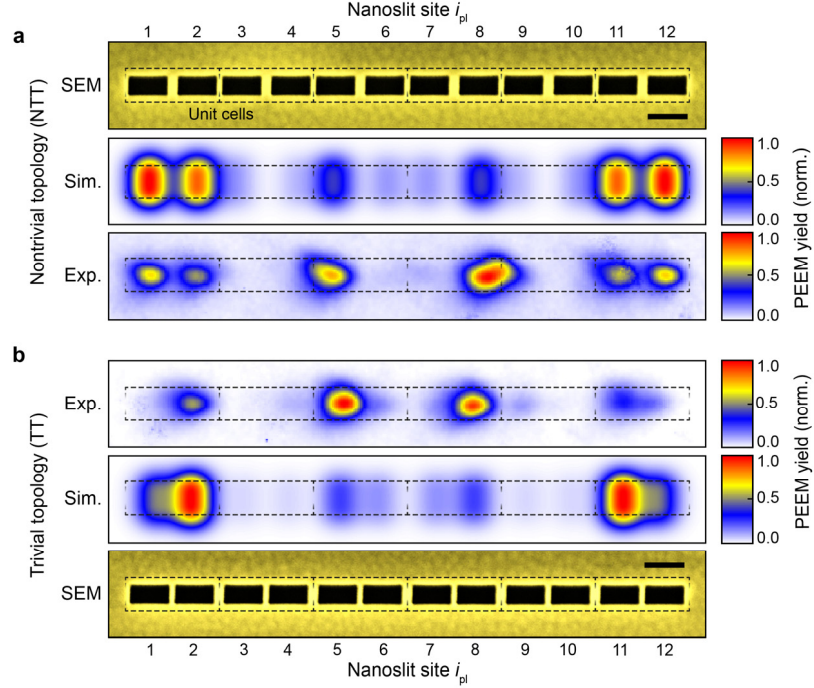

**Fig. 4: Imaging of the near-field intensity distribution of the plasmonic edge-state mode via photoemission electron microscopy (PEEM). a,** Plasmonic nanoslit chain in the nontrivial configuration ($b_1$ = 24 nm and $b_2$ = 12 nm). Top: SEM image. Simulated (middle) and measured (bottom) spatially resolved photoelectron emission yield upon pulsed laser irradiation (normal incidence, $\lambda_0 = 675\ nm \cong 1.837\ eV$). The nonlinearity of the photoemission process is *N* = 3 (see Supporting Information. The vectorial field data of FDTD simulations are acquired from a planar monitor, located 3 nm above the surface of the Au microplatelet, with a uniform mesh of 1 nm. See Methods for details. **b,** Plasmonic nanoslit chain in the trivial configuration ($b_1$ = 12 nm and $b_2$ = 24 nm). Measured (top) and simulated (middle) spatially resolved photoelectron emission yield upon pulsed laser irradiation at normal incidence. All scale bars: 100 nm. Dashed rectangles indicate the unit cells of the SSH chains. Bottom: SEM image.

Fig. 4a (Exp.) shows a typical PEEM image of an SSH chain that was fabricated using the parameters employed in the COMSOL simulations, Fig. 3c (nanoslit length $L = 100$ nm, width $w = 50$ nm, and bridge sizes $b_1 = 24$ nm and $b_2 = 12$ nm), i.e., the chain is expected to be in the NTT phase. This is also visualized by the dashed rectangles indicating the unit cells of the chain and revealing a larger intracell than intercell distance of the resonators. By comparing the SEM image (Fig. 4a, SEM) to the recorded PEEM yield map, the hot spots recorded in PEEM can be assigned to individual nanoslit resonators. It is found that the signal strength at the resonators varies over the chain. In particular, signals are very weak at the two central resonators $i_{\text{pl}} = 6$ and $i_{\text{pl}} = 7$. In contrast, there is strong photoelectron yield at the outermost resonators $i_{\text{pl}} = 1$ and $i_{\text{pl}} = 12$. The distribution of the photoelectron hot spots appears to be mirror-symmetric with respect to the chain center. Small differences in PEEM yield between resonators on mirrored positions can be explained by structural deviations that can lead to enhanced local fields, which in turn are translated into significant signal differences due to the non-

linearity of the photoemission process. In addition, small variations in the work function may play a role as well as residual cross-talk with nearby chains (minimum distance ~1.7 µm). Nevertheless, the characteristic mirror-symmetric hot-spot-like photoelectron emission pattern with maxima located at individual resonators suggests the excitation of a distinct collective plasmonic mode.

In Fig. 4b (Exp.) we show a typical PEEM image of a chain with the same nanoresonators ($L = 100$ nm and $w = 50$ nm) but a bridge-size sequence resulting in the TT phase ($b_1 = 12$ nm and $b_2 = 24$ nm). The indicated unit cells show a correspondingly larger intercell distance between the nanoresonators (Fig. 4b, SEM). As for the NTT chain, we observe hot spots in the emission of photoelectrons attributable to specific resonator positions as well as an overall mirror symmetry of that pattern. Strikingly, the hot spots on the two outermost resonators of the TT chain, i.e., $i_{\text{pl}} = 1$ and $i_{\text{pl}} = 12$, are very weak compared to the bulk hot spots of the same chain, while the hot spots of the outermost resonators of the NTT chain are clearly visible and stand out compared to the overall bulk signal. This indicates that the PEEM image of the NTT chain indeed reflects a mid-gap mode whose intensity is localized mainly at the two outer resonators.

To ensure that the visibility and brightness of the hot spots at the outer ends of the NTT chain compared to the TT chain are due to a plasmonic mode and not due to local defects in combination with the relatively large dynamic signal range associated with nonlinear photoelectron emission ($N = 3$ in this case, see Methods and Supporting Information section S5), we simulated the PEEM image signal $Y(\mathbf{r})$ using the vectorial local response function $\mathbf{R}(x, y, z, \omega)$ as retrieved from a planar monitor in FDTD simulations 3 nm above the Au microplatelet surface as well as the measured laser spectrum $E_L(\omega)$ (see Methods for details). The simulated PEEM images are displayed in Fig. 4a (Sim.) and Fig. 4b (Sim.) for the NTT and TT chain, respectively. The hot-spot patterns in the simulated PEEM images match the pattern in the measured images to a very large degree, in particular with respect to their position. The simulations confirm that the hot spots at the outermost nanoslit resonators $i_{\text{pl}} = 1$ and $i_{\text{pl}} = 12$ of the NTT chain, which are clearly visible in the PEEM experiment, can be assigned to the energetically higher lying mid-gap mode. This mid-gap mode, which matches the spectral window of the laser and is accessible from far field due to its even parity, can be unambiguously identified in the FDTD response function (see Supporting Information section S2). This kind of mode is not present in the FDTD response function of the TT chain and consequently only weak photoemission can be detected at the outer two nanoslit resonators of the TT chain, in both experiment and simulation.

Taking the typical structure of the eigenmode of the edge state (Fig. 3c, mode pattern at 1.810 eV) as the basis for the PEEM image contrast, however, the question arises as to why no alternating sequence of high and low photoemission signals from resonator to resonator can be observed in both the measured and simulated PEEM image of the NTT chain. The reason for this lies in the interplay of the topological invariant of the SSH chain, i.e., a geometric phase acquired over the Brillouin zone, which is called the Zak phase $\phi_{Zak}$[32], and the excitation of the entire chain with a homogeneous wavefront and polarization direction of the incident field. In the NTT configuration of the SSH chain, the Zak phase takes the value $\phi_{Zak} = \pi$, while $\phi_{Zak} = 0$ for the TT configuration. In our plasmonic system, $\phi_{Zak} = \pi$ signifies that the electromagnetic field distribution acquires a phase shift of $\pi$ when moving across the boundaries of the unit cells[14,33], indicated by the dashed rectangles in Fig. 4a. In other words, the electric field of the mid-gap mode changes its sign from resonator $i_{pl} = 1$ to $i_{pl} = 3$ to $i_{pl} = 5$, and so on, just as the real part of the edge-state wavefunctions in the quantum model changes sign at the same nanoslit sites (Fig. 3b). Consequently, an incident plane wave with a given polarization direction would interfere destructively with the edge-state mode field at $i_{pl} = 3$, for example, if the incident field interferes constructively with the edge-state mode field at nanoslit resonators $i_{pl} = 1$ and $i_{pl} = 5$ at the same time. Since widefield excitation is used in both experiment and simulation, the Zak phase of the NTT chain is indicated by the absence of the photoemission signal at $i_{pl} = 3$ and $i_{pl} = 10$ while at the same time the photoemission signal at the outer two nanoslit resonators $i_{pl} = 1$ and $i_{pl} = 12$ is clearly recognizable. In fact, the archetypal look of the edge-state mode as seen in Fig. 3c is obtained in FDTD simulations if only the outer two nanoslit resonators $i_{pl} = 1$ and $i_{pl} = 12$ are selectively excited by spatially-restricted plane-wave sources (see Supporting Information section S2). Therefore, considering the plane-wave excitation condition, the recorded hot-spot pattern of photoemission of the NTT chain is indeed the expected signature of a topological edge-state mode.

**Main: Conclusions**

We have fabricated high-precision plasmonic SSH chains with a unit cell length of 236 nm from individual nanoslit resonators and were able to demonstrate and visualize the existence of archetypal edge states in the nontrivial configuration of the chain experimentally via PEEM revealing strong spatial field localization at the outermost nanoslits as well as the characteristic mode pattern of a mid-gap state. This behavior is theoretically predicted by a quantum model considering nearest-neighbor coupling and a matching quasi-normal mode analysis. By combining He-FIB and monocrystalline Au microplatelets as the structural material, we precisely realized the required nanoslit spacings down to 12 nm, with a standard deviation smaller than 1 nm leading to coupling strengths between two nanoslit resonators in the ultra-strong coupling regime. This results in a band gap of 320 meV for the entire SSH chain, which corresponds to more than 15 % of the nanoslit resonance energy. The near-field mode pattern associated with this edge-state mode, as imaged by plasmon-assisted photoemission microscopy, has been explained considering the impact of the Zak phase $\phi_{\text{Zak}} = \pi$ in the topologically nontrivial phase and the far-field excitation conditions.

Our unambiguous experimental demonstration of the ability to control and spectrally isolate topological states in plasmonic systems, which exhibit subwavelength localization of plasmonic excitations in edge states, represents a decisive advancement in several ways. First, it paves the way for achieving strong coupling between single emitters and topologically protected edge states, enabling the exploration of fundamental questions about extending topological protection to hybrid quantum states of light and matter. Second, it opens new possibilities for the design of topological edge modes in 2D arrangements[18,19]. This will enable advanced functionalities such as unidirectional energy transport with minimal loss, even stronger energy confinement in corner states, and the exploitation of long-range coherence between different corner states.

**Methods**

**Photoemission electron microscopy (PEEM) setup**

PEEM images were acquired using a Yb-doped fiber laser (Amplitude Systèmes, Tangerine HP, 1030 nm, 35 µJ, ~320 fs, 1 MHz repetition rate), which pumps a two-branch noncollinear optical parametric amplifier (NOPA, Riedle group, LMU Munich) achieving a spectral range of $\Delta\lambda$ = 120 nm (full width at tenth maximum, corresponding to $\Delta E$ = 200 meV) at the tailored central wavelength of $\lambda_L$ = 675 nm (or $E_L$ = 1.837 eV). The pulses were compressed with a prism compressor and guided through a liquid-crystal-display-based spatial light modulator (Jenoptik, SLM-S640d USB) in a 4f geometry, enabling double-pulse scans. Pulse characterization was performed using spectral phase interferometry for direct electric-field reconstruction setup (FC Spider VIS, APE Angewandte Physik und Elektronik GmbH).

The sample was irradiated under normal incidence with pulses of ~20 fs duration, a beam radius of $r_x$ = 270 µm and $r_y$ = 405 µm, and 46 nJ pulse energy, preventing space-charge effects. The polarization of the laser pulses was set perpendicular to the SSH chains. Active beam stabilization (TEM Messtechnik, Aligna 4D) was used to correct for vibrational instabilities between laser setup and the separate PEEM table.

Photoelectrons emitted from a 5 µm diameter field of view were collected using an aberration-corrected photoemission electron microscope (AC-LEEM III, Elmitec Elektronenmikroskopie GmbH), filtered in k-space via a 60 µm contrast aperture, and transferred to the detection unit. The detection unit consisted of two chevron-type microchannel plates (MCP) for electron multiplication, a phosphor screen for electron-to-photon conversion, and a charge-coupled device (CCD) camera for image acquisition.

We collected and analyzed both static images and double-pulse scans ($\tau$ = 0 to 81 fs with a step size of $\delta\tau$ = 0.3 fs) with an integration time of 90 s. The images from the double-pulse scan were integrated to obtain the static response after applying drift correction to compensate for small spatial drifts over the ~15 h measurement time. Drift correction was performed using reference images acquired with single-pulse excitation and an integration time of 30 s. Through this procedure we receive PEEM images with improved signal-to-noise ratio but the same mode pattern as in the static images. Further details on the experimental setup can be found elsewhere[29].

**Modelling PEEM yield with FDTD data**

FDTD simulations considered a 700 nm thick glass substrate and a 38 nm thick Au layer describing a microplatelet. Material data were taken from Palik's handbook[34] as well as from Johnson and Christy[35], respectively. The nanoslit resonators were modelled as rectangular holes with rounded corners and edges that exhibit a 6 nm radius of curvature. To take the He-FIB process into account, the nanoslit resonators extended 10 nm beyond the Au layer into the glass. The SSH chain was discretized using a finer mesh with a spatial resolution of 1 nm. To take the far-field excitation conditions into account, the SSH chain was homogeneously excited by plane-wave total-field scattered-field source, with a polarization perpendicular to the chain.

The FDTD response function $\mathbf{R}(x,y,z,\omega)$ determined by these excitation conditions, i.e., the superposition of incoming field and excited plasmonic near-field, was recorded in a planar monitor 3 nm above the Au surface. We obtained the local near-field responsible for photoemission via $\mathbf{E}_{\text{loc}}(x,y,z,t) = \text{FT}\{\mathbf{R}(x,y,z,\omega) \cdot E_{\text{L}}(\omega)\}$, where $E_{\text{L}}(\omega)$ is the spectral amplitude of the measured laser spectrum and $\text{FT}\{...\}$ denotes the Fourier transform. For modeling the PEEM yield according to the ansatz of the surface and volume photoelectric effect, we used the multi-plasmon photoemission model of Podbiel and co-workers[36] in which the local field is split into a component perpendicular, i.e., $E_{\text{loc}}^{\perp}(x,y,z,t) = |E_{z,\text{loc}}(x,y,z,t)|$, and parallel, i.e., $E_{\text{loc}}^{\parallel}(x,y,z,t) = \sqrt{|E_{x,\text{loc}}(x,y,z,t)|^2 + |E_{y,\text{loc}}(x,y,z,t)|^2}$, to the surface such that

$$Y(x,y,z) \propto \int_{-\infty}^{\infty} \left[ (E_{\text{loc}}^{\parallel}(x,y,z,t))^2 + \alpha^2 (E_{\text{loc}}^{\perp}(x,y,z,t))^2 \right]^{2N} dt,$$

where $N = 3$ is the number of absorbed photons, as determined experimentally (see Supporting Information section S5). The value of $\alpha = 5.8$ is an empirically found value[36] that usually is associated with a specific facet of the Au single crystal but also works reasonably well in our case, although it is not known which facets contribute to photoemission here. To model the PEEM images in Fig. 4, we evaluated $Y(x,y,z)$ only at the points where gold is present as material in the lateral structural profile of the SSH chain. Finally, we convoluted $Y(x,y,z)$ with a 2D Gaussian that exhibits a FWHM of 70 nm to account for the reduced resolution of the photoemission hot spots near the nanostructured surfaces and any imaging errors in the PEEM alignment.

**Eigenmode decomposition with COMSOL**

The structures and material properties used in our COMSOL simulations are identical to those in the FDTD simulations. A perfectly matched layer with a thickness of 250 nm was applied to suppress backward reflections at all boundaries of the simulation volume. Due to the mirror symmetry of the SSH chain structure with respect to the *xz* and *yz* planes, a refined mesh was required for only one quarter of the entire structure and the fields retrieved by mode analysis were accordingly mapped to the residual three domains. An eigenfrequency solver was employed to search for all possible states within the system around the plasmonic nanoslit resonance, while filtering out any unphysical modes caused by numerical artifacts. Finally, after post-processing, we normalized all mode patterns to the maximum electric field intensity.

**Data availability**

The data associated with this study, including raw FDTD and COMSOL files, is freely available in WueData at [link will be inserted upon publication], reference number [Ref].

**Acknowledgements**

We gratefully acknowledge funding by the Deutsche Forschungsgemeinschaft (DFG, German Research Foundation) under Germany's Excellence Strategy through the Würzburg-Dresden Cluster of Excellence on Complexity and Topology in Quantum Matter, ct.qmat (EXC 2147, Project ID ST0462019) (B.H. & R.T.) as well as through a DFG project (INST 93/959-1 FUGG) (B.H.). This works was also supported by the "Staatsministerium für Wissenschaft und Kunst" of the state of Bavaria within the framework "Hightech Agenda Bayern Plus", specifically, the project "Integrated Spin Systems for Quantum Sensors (IQ-Sense)" which is part of the "Munich Quantum Valley" (B.H.). We also acknowledge the QuantERA II Programme that has received funding from the European Union's Horizon 2020 research and innovation programme under Grant Agreement No 101017733, and with DFG (HE5618/12-1), SFI (Ireland), and NCN (Poland). All authors thank Dr. Thorsten Feichtner for providing the Blender image in Fig. 1a.

## Author Contributions

## Competing interests

The authors declare no competing interests.

# Supporting Information

## Plasmonic Su–Schrieffer–Heeger chains with strong coupling amplitudes

*Benedikt Schurr[1,*], Matthias Hensen[2,*], Luisa Brenneis[2], Philipp Kessler[2], Jin Qin[1], Victor Lisinetskii[2], Ronny Thomale[3], Tobias Brixner[2,†], and Bert Hecht[1,†]*

[1] NanoOptics & Biophotonics Group, Experimental Physics 5, University of Würzburg, Am Hubland, 97074 Würzburg, Germany

[2] Institut für Physikalische und Theoretische Chemie, Universität Würzburg, Am Hubland, 97074 Würzburg, Germany

[3] Institute for Theoretical Physics and Astrophysics, University of Würzburg, 97074 Würzburg, Germany

*Shared first authorship

†Corresponding authors:
bert.hecht@uni-wuerzburg.de (ORCID 0000-0002-4883-8676);
tobias.brixner@uni-wuerzburg.de (ORCID 0000-0002-6529-704X);

# Contents

## S1 Fabrication of nanoslit SSH chains

**S1.1 Sample layout and determination of the thickness of the gold microplatelet**

All Su–Schrieffer–Heeger (SSH) chains discussed in the manuscript were fabricated on a hole mask sample (Fig. S1.1a), specifically designed for photoemission electron microscopy (PEEM) experiments. For this, a 24 mm × 24 mm microscopy glass cover slip (cover glasses, Menzel Gläser) was used as a glass substrate which exhibits a thickness of 0.17 mm. Before processing, the glass substrate was thoroughly cleaned using a multi-step procedure. First, the glass substrate was placed in an ultrasonic bath with ultra-pure acetone and ethanol for 10–15 min. Afterwards, the glass substrate was rinsed by ultrapure water and blow-dried by pressurized nitrogen gas. Finally, remaining residues were removed by plasma cleaning (PlasmaFlecto 10, Plasma Technology) for 5 min at 250 W with an oxygen flow of 5 standard cubic centimeters per minute (sccm). Next, the glass substrates underwent an optical lithography process, during which they were coated with a 15 nm chromium layer followed by approximately 70 nm of thermally evaporated gold. An optical mask was used to obtain the desired patterns, e.g., marker structures, like a coordinate system. Specific sample positions were encoded by specific hole pattern matrices (see Fig. S1.1a, b). The circular shape of the holes prevents electrical charging at sharp corners during PEEM experiments. Furthermore, larger circular hole structures with diameters of up to 100 μm were fabricated on the sample using optical lithography. These large holes served as deposition sites for the monocrystalline gold microplatelets, into which the SSH chains were later written (Fig. S1.1c).

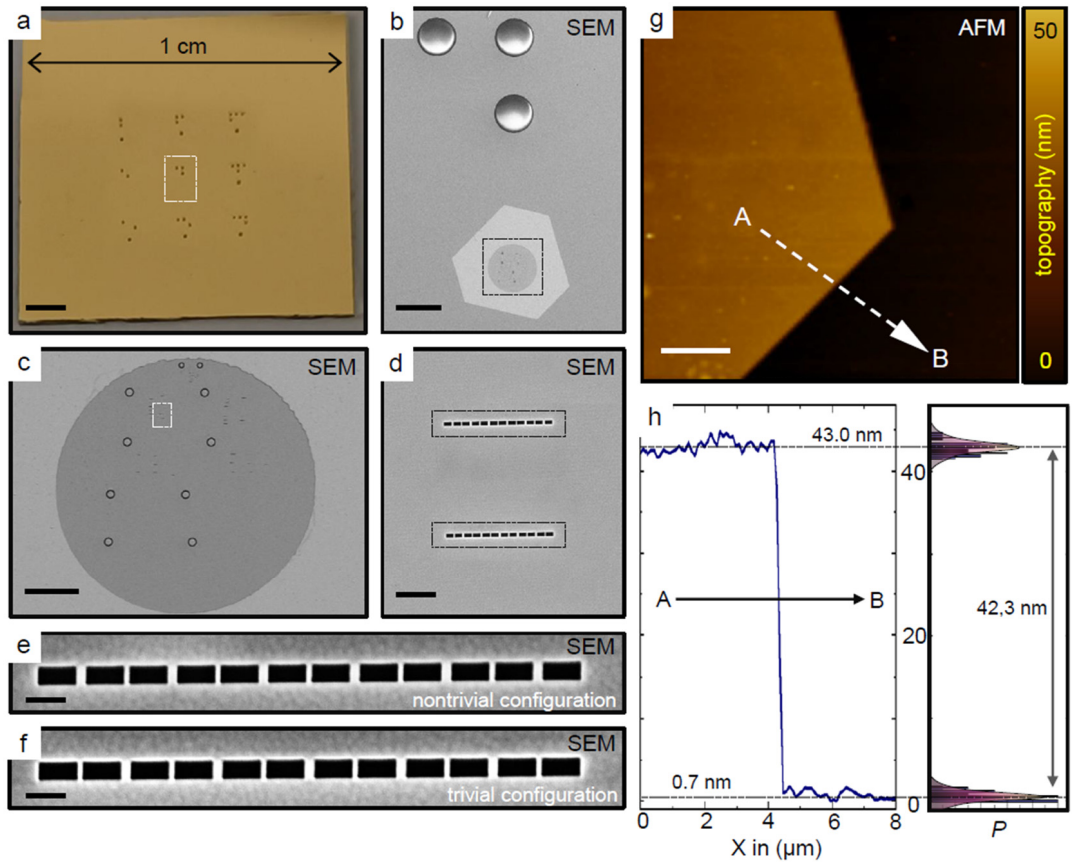

**Fig. S1.1: Sample layout and gold microplatelet thickness**. **a,** Camera image of the sample used for wide-field PEEM experiments. Since a mask was used before the evaporation process, a specific 3×3 hole pattern matrix was created on the sample, providing circular holes with diameters of up to 100 µm for the microplatelet transfer. A monocrystalline gold microplatelet was placed on such a hole to provide a monocrystalline gold layer on top of the pure glass substrate for optical measurements. The scale bar is 1 mm. **b,** SEM scan of one matrix spot on the sample, (marked in **a**) showing a positioned Au microplatelet on a prepared hole. The scale bar is 100 µm. **c,** Zoom-in SEM scan of the marker area in **b**. Marker structures (ring-like objects) were cut into the 2D monocrystalline gold using Ga FIB milling. These structures, measuring up to 5 µm, were used to locate the position of the SSH chains in the PEEM. The scale bar is 20 µm. **d,** SEM image of a nontrivial and trivial SSH chain, as a zoom-in of the marked position in **c** next to a marker. The scale bar is 500 nm. **e** and **f,** Close-up SEM scans of the nontrivial **(e)** and trivial **(f)** SSH chain configurations. The scale bars are each 100 nm. **g,** AFM scan of a corner of the used monocrystalline Au microplatelet used in **b–f**. The scale bar is 1 µm. **h,** Analysis of the height profile from the line cut shown in **g**. The thickness of the microplatelet was estimated from the histogram to be 42.3 nm.

The synthesis of monocrystalline gold microplatelets with high aspect ratios is described in detail in Wu et al.[1] In the next step, an Au microplatelet was transferred from the growing substrate onto the hole mask sample, using an in-house developed technique by covering a microplatelet with a droplet of polymethylmethacrylate (PMMA). After baking the PMMA microplatelet composite at 100 °C for 1 h, the compound was picked up from the growing sample and placed onto the desired position on the hole mask. The PMMA was finally removed using an acetone bath.

After the transfer, the sample was again plasma-cleaned with the same specifications as before. In a final step, the hole mask sample was cut to a size of 1 cm × 1 cm using a diamond cutter tool to fit the sample holder of the PEEM.

The monocrystalline gold microplatelets are particularly suitable for nanofabrication using focused-ion-beam (FIB) milling. Here, we have used a helium ion microscope (Orion NanoFab, Zeiss) which combines a gallium (Ga) and a helium (He) FIB. While the Ga-FIB is commonly used to fabricate rather coarse and larger structures on short time scales, the He FIB can be used for fine structuring objects with nm precision. Therefore, Ga-FIB was used to create ring-like marker structures inside the microplatelet (Fig. S1.1c) to facilitate finding the desired position of the SSH-chains with PEEM. while to fabricate sub–10 nm nanostructures[2]. For the fabrication of nanoslits and SSH chains (Fig. S1.1d–f), the He-FIB was focused down to a spot size of below 1 nm and an acceleration voltage of 35 kV was used, providing an average ion current of 3.2–3.4 pA.

For the correct modelling of nanoslits and SSH chains, as well as for finding suited patterning parameters for the fabrication procedure, the thickness of the gold microplatelet, i.e., the depth of the individual nanoslits, plays an essential role. Atomic force microscopy (AFM) was used to measure the thickness of the microplatelet, using a line-cut as depicted in Fig. S1.1g. As the AFM scan was performed at the very edge of the microplatelet, the maximum thickness of the unstructured microplatelet is determined. The height profile and the corresponding histogram provide a thickness of the microplatelet of $(42.3 \pm 0.9)$ nm (Fig. S1.1h). Note that this value is valid at the edge of the microplatelet but, most likely, it is slightly smaller (a few nm) in the area where the fabrication is performed. This is due to the permanent removal of gold layers while scanning over the area of interest to find the preselected area. To reduce this effect, the number of He-FIB scans was reduced to a minimum before fabricating the final SSH chains.

**S1.2 Fabrication of nanoslit resonators using high-precision He FIB milling**

In this section, the fabrication of the individual nanoslits and SSH chains is explained. For the fabrication of single nanoslits, rectangular FIB patterns were designed, and the corresponding FIB parameters were adjusted. Since the amount of gold milled from the microplatelet is highly dependent on the energy of the He ions, the ion current (rate), and the total sputtering duration, these parameters were carefully adjusted to achieve a precise and uniform cut of the predesigned rectangle into gold. Here, we have used a 35 kV He FIB with an average current of 3.4 pA that was focused onto the sample's surface. The decisive property of the FIB process is the He ion dose that is applied to the sample. Several parameters like the dwell time of the He-beam, or the number of FIB repeats determine the total dose. To determine the minimum dose required to completely cut through an approximately 40 nm thick gold microplatelet, dose tests were conducted. In Fig. S1.2a, a 3D AFM scan of a 3×3 nanoslit matrix is presented, showing different sputtering results for nine different dose values. A clear trend can be observed between a higher dose and a deeper incision into the gold. The doses for this test were varied between 0.1 C/cm² and 1.6 C/cm².

For a quantitative analysis of the relation between dose and FIB depth, the measured nanoslit depths are plotted versus the applied doses (Fig. S1.2b). A linear relationship is observed, showing an increase in nanoslit depth as a function of dose, with a rate of $-53.8\ (\mathrm{nm}\cdot\mathrm{cm}^2)/\mathrm{C}$. Furthermore, the last two data points corresponding to doses of 1.4 C/cm² and 1.6 C/cm², respectively, indicate a saturation of the achievable depth, which is attributed to the limited penetration depth of the AFM tip exhibiting a tip radius of ~10 nm. This is why the last two data points have been excluded from the evaluation. The insets in Fig. S1.2b depicts line cuts (labelled as A and B) from the AFM scan in Fig. S1.2a. While the bottom one shows a cut providing the width of the nanoslit, the inset at the top displays a line cut along three different nanoslits, fabricated with different He ion doses.

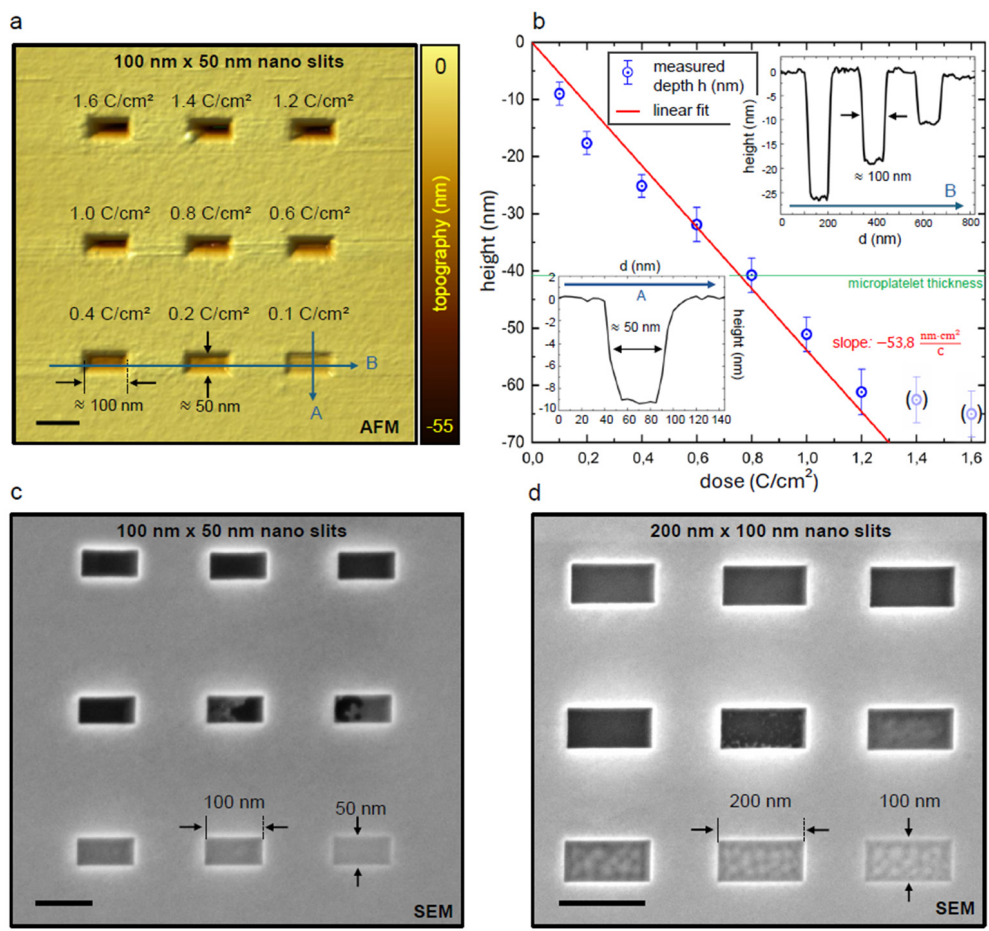

**Fig. S1.2: Nanoslit fabrication and dose tests. a,** 3D AFM scan of nine helium-ion-milled rectangular nanoslits, each measuring 50 nm in width and 100 nm in length. To check the depth evolution of the fabricated nanoslits inside the Au microplatelet with increasing He ion doses, nine nanoslits have been milled out using doses ranging from 0.1 C/cm² to 1.6 C/cm². **b,** Measured nanoslit depths of the dose test structures from **a**. Increasing He ion doses lead to larger depths of the created nanoslit. As displayed in **b**, a linear behavior can be found, exhibiting a slope of the depth-dose relation of approximately -53.8 (nm · cm²)/C. A saturation behavior above 1.2 C/cm² (excluded data) occurs due to a limited resolution of the AFM with respect to the AFM tip radius. **c,** SEM image of the nine nanoslits presented in **a**. As visible from the contrasts, the dose-depth relation found in **a, b** could also be found in the SEM image. **d,** SEM scan of nanoslits twice as large compared to those in a-c, providing the same aspect ratio of 2:1.

This line cut clearly shows that higher He ion doses lead to deeper nanoslits. Based on the dose tests and the fact that the gold microplatelet exhibits a maximum thickness of approximately 41 nm, a suitable dose for fabrication was determined to be between 1.0 C/cm² and 1.1 C/cm². This determined value ensures that the microplatelet is completely cut through while minimizing milling into the glass substrate. From Fig. S1.2b, the depth of milling into the glass was estimated to range between 12 nm and 15 nm. As previously discussed, the area of interest was scanned two to three times before fabrication for orientation and sample adjustments, so that a reduction of microplatelet thickness by a few nm is expected.

In Fig. S1.2c, a scanning electron microscope (SEM) scan of the same 3×3 nanoslit matrix from Fig. S1.2a is displayed. The overall trend and observations from the AFM investigation are confirmed.

For low doses, a very low contrast is observed between the nanoslits and the surrounding, unstructured gold. On the one hand, for doses just below 1.0 C/cm² the SEM scan clearly shows that the He beam has nearly cut through the microplatelet, while some remaining gold can be observed inside the nanoslit. On the other hand, doses of 1.0 C/cm² and above indicate a full cut through the gold layers and an increasing sputtering of the glass substrate beneath. It should be noted that in the SEM scan, no significant differences could be observed for the nanoslits fabricated with He FIB doses of 1.2 C/cm² and higher since allegedly the contrast, generated by the scattered electrons at the glass substrate inside the nanoslit, remains approximately unchanged for doses ranging from 1.2 C/cm² to 1.6 C/cm². In Fig. S1.2d, another SEM scan of a 3×3 nanoslit matrix is depicted verifying the observations from the 50 nm × 100 nm nanoslits. In this case, the nanoslits were chosen to provide twice the aspect ratio of the nanoslits used for the actual SSH chains.

For the actual fabrication of the SSH chains, FIB patterns were designed to create a chain of twelve identical nanoslits, aligned in a row, and separated by nanoscale bridges of alternating bridge sizes $b_1$ and $b_2$. The widths of the bridges were designed to be 12 nm for the small bridges and 24 nm for the large ones. Since a one-by-one fabrication of the individual nanoslits in an SSH chain inevitably led to the destruction of bridges, it was essential to change the fabrication technique towards a "parallel-patterning approach". In this process, the individual nanoslits were produced simultaneously. Although the nanoslits were removed layer-by-layer, i.e., sequentially, the process restarted always at the same chain end after each pass, ensuring uniform material removal. Using this technique, it was possible to fabricate the delicate bridges between individual nanoslits.

**S1.3 Evaluation of bridge sizes of nanoslit SSH chains**

The nanoscale bridges $b_1$ and $b_2$ separating the nanoslit resonators are the key parameters that determine the coupling between individual nanoslits. It is therefore important to precisely design, fabricate, and measure the widths of these bridges.

In Fig. S1.3a, an SEM scan of a nontrivial SSH chain is depicted (top panel), in which all bridges are labeled. Every fabricated SSH chain exhibits eleven bridges since the chains consist of twelve nanoslit resonators. For the case of nontrivial topology (NTT), the chain has six large bridges of width $b_1$ and five narrow bridges of width $b_2$. A statistical evaluation of all bridge widths of an NTT SSH chain was performed and bridge widths were measured directly after the fabrication process using SEM (GeminiSEM 450, Zeiss). In the bottom panel of Fig. S1.3a, a zoomed-in version of the image of the complete chain in the top-panel is shown. To quantify the bridge widths each bridge was measured along three different line-cuts, as indicated by the three blue lines.

The resulting bridge widths are plotted and analyzed in Fig. S1.3b. As is visible from the distribution of the data points, the bridge widths deviate only slightly from the designed bridge width of 24 nm and 12 nm. For the large bridges (black triangles), a mean value with an associated standard deviation of $(24.8 \pm 0.9)$ **nm** is found, while for the small bridges (red dots), the mean value is found to be $(12.7 \pm 0.5)$ **nm**. The $1\sigma$ standard deviation is indicated in the plot with gray and light-red rectangular stripes. This evaluation suggests that the fabrication precision is close to, or even below, 1 nm. In this fabrication approach, the bridge widths $b_1$ and $b_2$ are both slightly larger compared to the target values. This, however, does not influence the overall topological behavior of the SSH chains and only slightly modifies the couplings.

In Fig. S1.3c,d, we show two exemplary line cuts for the bridge width analysis. As indicated in the plots, the bridge width was determined by calculating the full width at half maximum (FWHM) for each line cut.

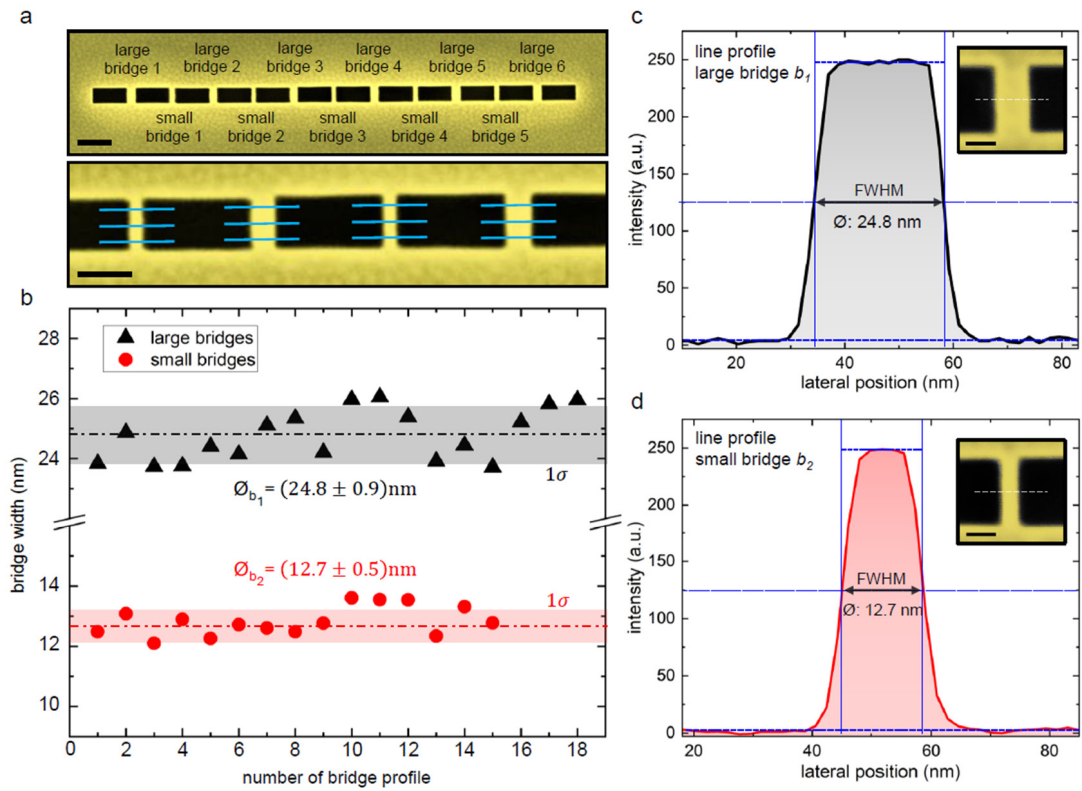

**Fig. S1.3: Evaluation of bridge widths of an SSH chain.** **a**, SEM scan of a complete NTT SSH chain with labeled bridges (top panel). To evaluate the large and small bridge widths of $b_1$ and $b_2$, respectively, three contrast line cuts have been extracted from a zoom-in SEM image (bottom panel) for each bridge to obtain average values. **b,** Measured width values for the smaller (red dots) and larger bridges (black triangles). Both the large and small bridge widths exhibit a narrow distribution around their mean values of $(24.8 \pm 0.9)$ nm and $(12.7 \pm 0.5)$ nm, respectively. Additionally, the 1σ interval is displayed for both measurement series. **c** and **d**, exemplary line cuts for the analysis of the bridge widths for the large bridges $b_1$ (**c**) and for the small bridges $b_2$ (**d**). The values of the bridge widths were determined by the full width at half maximum (FWHM) of each of the line cuts. The insets in **c** and **d** show close-up SEM scans of the corresponding bridge $b_1$ and $b_2$.

## S2 Zak phase and its impact on near-field patterns

### S2.1 Zak phase influence on the electric field distribution of edge-state eigenmodes

In the main manuscript, we discuss that the absence of PEEM yield at nanoslit resonators $i_{pl} = 3$ and $i_{pl} = 10$ is due to the combination of the Zak phase and wide-field excitation conditions. To demonstrate that the Zak phase is present at all in the NTT configuration of the plasmonic SSH chain, we first show in Fig. S2.1 the $E_y$ component of the two mid-gap modes as obtained from our COMSOL eigenmode decomposition. We show here the $E_y$ component because the lateral field component perpendicular to the SSH chain corresponds to the resonant excitation condition of the nanoslit resonators. Indeed, for the mid-gap mode with even parity (Fig. S2.1a) the field amplitude at nanoslit resonators $i_{pl} = 1,3,5$ exhibits an alternating sign, while the amplitude decreases into the bulk. According to the Zak phase of π, i.e., the change of signs in field amplitude across the unit cell, nanoslit resonators $i_{pl} = 2$ and $i_{pl} = 4$ exhibit electric field strength close to zero. Starting from the right end of the chain, the same amplitude progression is observable due to even parity. Note that one reason for a non-zero field amplitude at nanoslit resonators $i_{pl} = 2,4,6$ and $i_{pl} = 11,9,7$ are retardation effects in our finite-sized structures[3]. Additionally, the total number of resonators, i.e., twelve, is so small that there is a significant overlap between the edge state at the left end of the chain and the edge state at the right end of the chain so that even within the quasi-static limit the mid-gap modes, resulting from hybridization of the two edge states, would also show a non-zero field amplitude at the respective nanoslit resonators. The odd-parity mid-gap mode (Fig. S2.1b) shows the same behavior in the electric field distribution caused by the Zak phase, except that the signs of the fields on the right side of the chain are inverted with respect to the left side due to the odd parity.

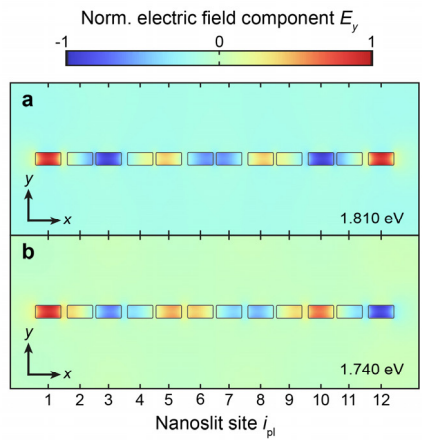

**Fig. S2.1: Real-valued $E_y$ component of the two plasmonic mid-gap modes retrieved from COMSOL simulations. a,** Even-parity mid-gap mode. **b,** Odd-parity mid-gap mode. The field strength is normalized with respect to each specific eigenmode, and the near-field distributions are retrieved at half of the nanoslit height.

**S2.2 Near-field patterns of the NTT SSH chain under different excitation conditions**

Now that we have shown that the Zak phase of π, archetypal for the nontrivial phase of the SSH chain, is present in our plasmonic system, we consider the response function of the NTT chain under wide-field excitation, as it corresponds to the excitation conditions of the PEEM experiment. Our assumption is that the spatially alternating sign of the field strength within the mid-gap mode leads to both constructive and destructive interference with the homogeneous field polarization of the excitation source, and thus that the photoemission yield does not reproduce the near-field distribution of the respective eigenmode one-to-one.

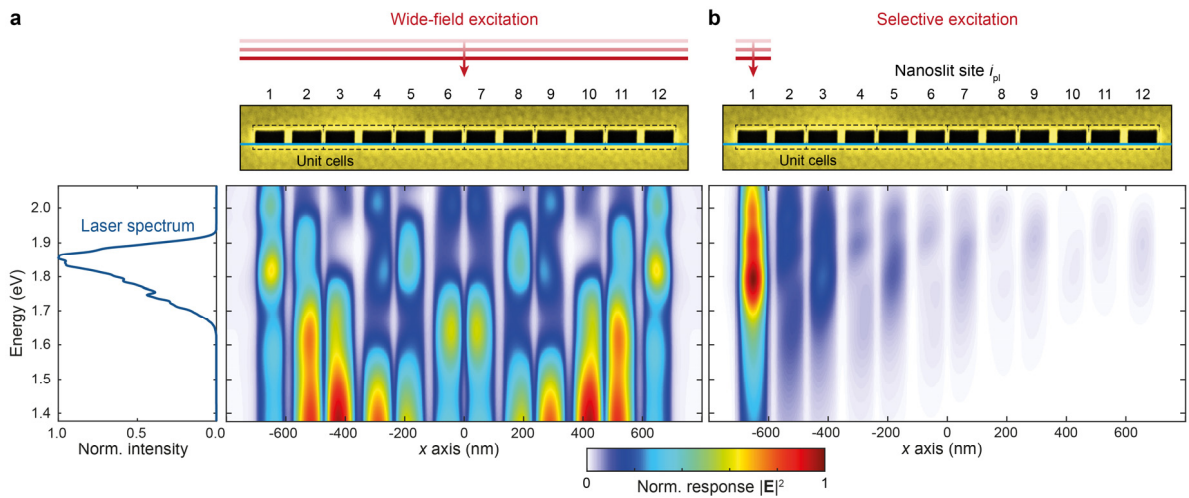

**Fig. S2.2: Spatially resolved absolute square of the FDTD response function for different excitation conditions.** **a**, Wide-field excitation with a plane-wave source, i.e., a total-field scattered-field (TFSF) source in Lumerical FDTD, spanning the entire chain. Additionally, the experimental laser spectrum is shown on the left. **b**, Selective excitation of the outermost nanoslit resonator on the left side by spatially restricting the TFSF plane-wave source to that specific resonator. Data is collected along the light-blue line, 3 nm above the gold surface.

In Fig. 2.2a , we show the absolute square of the FDTD-retrieved response function under wide-field excitation along the chain axis, close to the edge of the nanoslit resonators as indicated by the horizontal light-blue line in the top panel. Even if this quantity, here called "intensity" for the sake of simplicity, cannot be taken directly as a measure for the photoemission yield (see the model for calculating the PEEM yield in the main manuscript), it does provide a qualitative impression of the photoemission yield. First, the high field intensity at the outermost nanoslit resonators, at an excitation energy of ~1.8 eV, clearly shows that the even-parity mid-gap mode lies well within the laser spectrum. But in contrast to the pure eigenmode (Fig. 2.1a), the field intensity at nanoslit resonators $i_{pl} = 3$ and $i_{pl} = 10$ is close to zero, which is consistent with the low photoemission yield in the experiment (Fig. 4a, main manuscript). We attribute this behavior to the change of sign in the electric field from the outermost resonators $i_{pl} = 1$ and $i_{pl} = 12$ to resonators $i_{pl} = 3$ and $i_{pl} = 10$, respectively. According to the eigenmode decomposition in Fig. S2.1, the electric fields at $i_{pl} = 1$ and $i_{pl} = 5$ as well as $i_{pl} = 12$ and $i_{pl} = 8$ exhibit the same sign, which should result in a constructive interference of the incoming laser field and the near field of the mid-gap mode at $i_{pl} = 5$ and $i_{pl} = 8$. Indeed, looking the FDTD response in Fig. S2.2a at an excitation energy of ~1.8 eV, there is significant intensity at $i_{pl} = 5$ and $i_{pl} = 8$ that contributes to photoemission, as shown by the experimentally measured data as well as by the yield simulations (Fig. 4a, main manuscript).

For the sake of completeness, we also show the effect of selective excitation of the outermost nanoslit resonator $i_{pl} = 1$. As can be seen in Fig. S2.2 (right), the intensity decreases from nanoslit resonator $i_{pl} = 1$ via $i_{pl} = 3$ to $i_{pl} = 5$ and then further into the bulk, while the nanoslit resonators $i_{pl} = 2$ and $i_{pl} = 4$ in between show an abrupt drop in intensity. This resembles the behavior of mid-gap modes. However, a selective excitation of a single resonator does not allow to discriminate between the excitation of the even- and odd-parity mid-gap mode. Consequently, both are excited simultaneously. Since they add up constructively on the left side of the chain there is, due to the opposite parity, hardly any intensity on the right side of the SSH chain. Although they are expected to be spectrally separated by a few tens of meV, they still exhibit sufficient spectral overlap to cancel out themselves on the right side of the chain. The superposition of both mid-gap modes also explains why the spectral intensity at $i_{pl} = 1$ is shifted to lower energies compared to the spectral intensity at the same resonator under wide-field excitation conditions. The reason is that the energetically lower lying odd-parity mid-gap mode pulls the spectral weight to lower energies when both mid-gap modes are excited simultaneously.

In summary, it can be concluded that the dip in the measured and simulated photoemission yield at nanoslit resonators $i_{pl} = 3$ and $i_{pl} = 10$ is due to the interference of incident laser light and the even-parity mid-gap mode near-field distribution as determined by the Zak phase.

## S3 Eigenstates of the excitonic SSH chain in NTT and TT configuration

Here, we show in Fig. S3.1 and Fig. S3.2 the real part of the site-resolved wave functions of the excitonic SSH chains in nontrivial (NTT, left) and trivial (TT, right) configuration, divided into the six energetically lower and the six energetically higher states, respectively. The wave functions were retrieved by diagonalizing equation (1) in the main manuscript. In this representation, the alternating parity from eigenstate to eigenstate is clearly visible. We point out once again that due to the short chain length of twelve two-level systems, there is a wave function overlap of edge states which are mainly localized at the left and the right chain ends. Hybridization of these edge states results in the odd- and even-parity mid-gap states at $E = 1.840$ eV and $E = 1.878$ eV, respectively, in the NTT configuration.

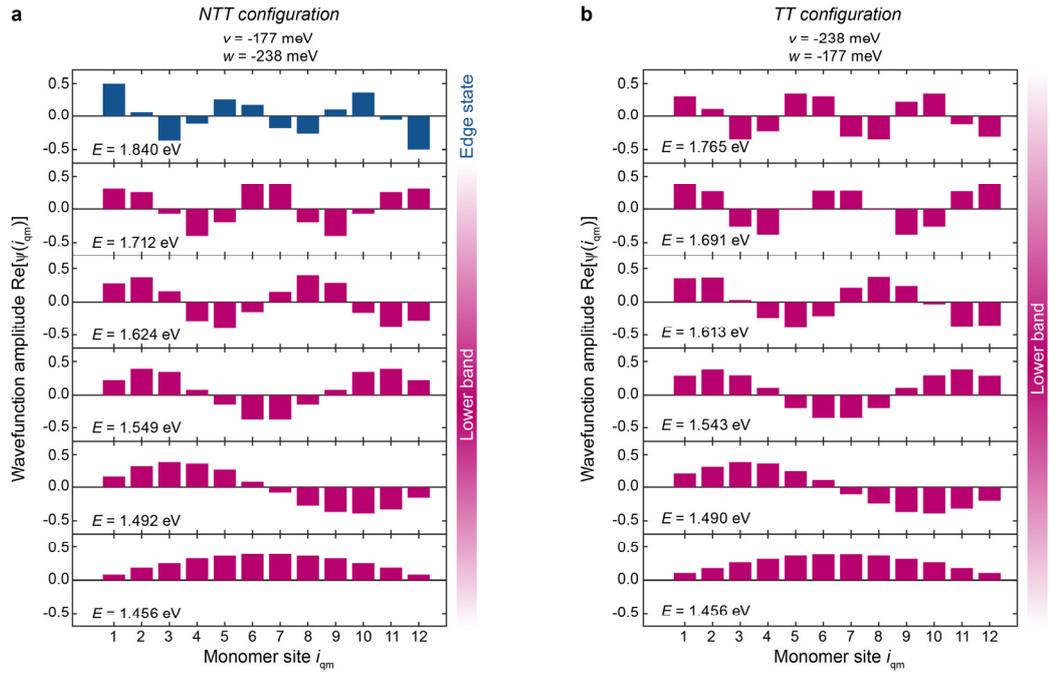

**Fig. S3.1: Site-resolved lower-band and edge-state wave functions (real part) of the excitonic SSH chain.** **a,** Nontrivial (NTT) chain configuration. **b,** Trivial (TT) chain configuration. Derived by diagonalization of equation (1) in the main manuscript.

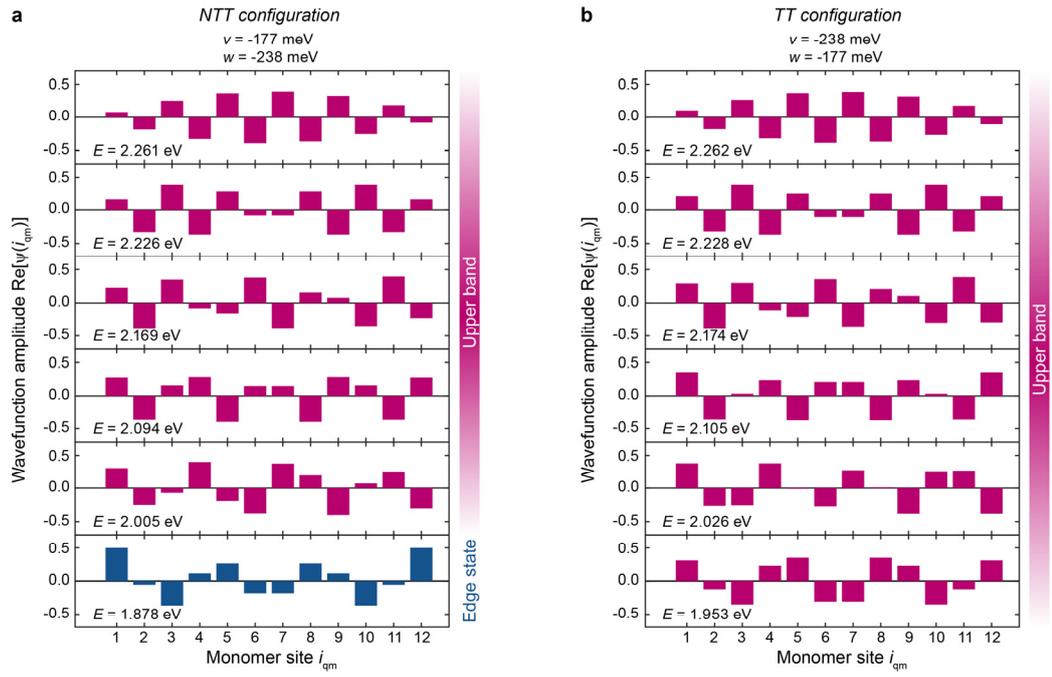

**Fig. S3.2: Site-resolved edge-state and upper-band wave functions (real part) of the excitonic SSH chain.** **a**, Nontrivial (NTT) chain configuration. **b**, Trivial (TT) chain configuration. Derived by diagonalization of equation (1) in the main manuscript.

# S4 Eigenmodes of the plasmonic SSH chain in NTT configuration

For the sake of completeness, we show all mode patterns (intensity) of the COMSOL eigenmode decomposition of the plasmonic SSH chain in NTT configuration in Fig. S4.1.

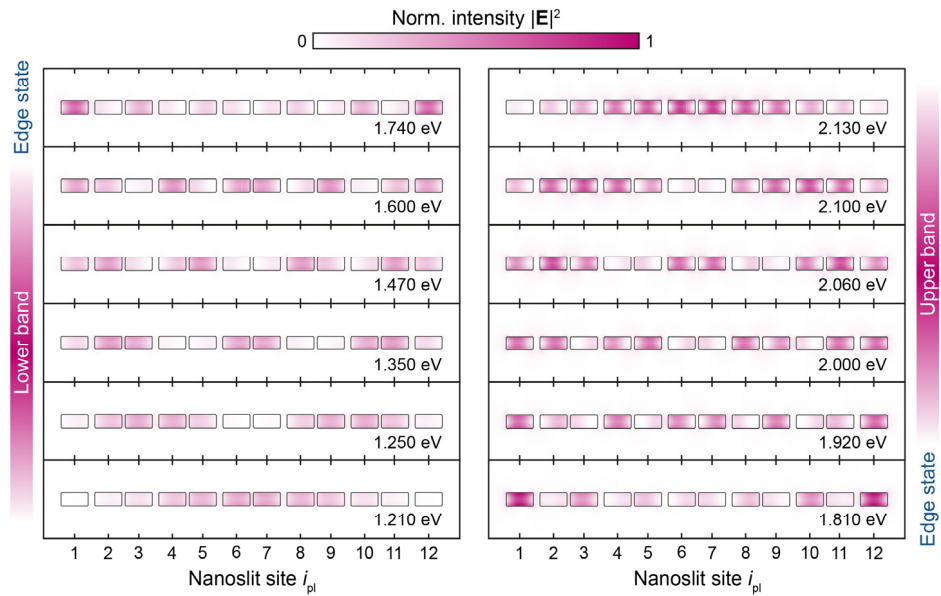

**Fig. S4.1: Intensity distribution of the plasmonic SSH chain eigenmodes for the nontrivial (NTT) configuration retrieved from a COMSOL eigenmode decomposition.** All mode patterns are normalized to a global maximum. Data is retrieved from a lateral monitor located at half the height of the nanoslits. The bridge sizes are $b_1$ = 24 nm and $b_2$ = 12 nm.

## S5 Determining the nonlinear order in photoemission

The dependence of the photoemission process on electric field polarization and its nonlinearity were investigated by measuring the PEEM yield as a function of the electric field polarization orientation. Figure S5.1 shows the spatially integrated, background-subtracted, and normalized PEEM yield of a nanoslit SSH chain with a nontrivial (a) and a trivial configuration (b). The angle $\alpha$ defines the relative orientation between the linear polarization of the excitation pulse(s) and the nanoslit chain. For $\alpha = 0°$ and $\alpha = 180°$ the polarization is aligned parallel to the long axis of the nanoslits and the chain, whereas at $\alpha = 90°$ (dashed line) the polarization is oriented parallel to the short axis of each nanoslit. The spatially integrated PEEM yield of both chains reaches a maximum at $\alpha = 90°$, consistent with the required electric field polarization along the short axis of the nanoslits for resonant excitation.

Due to the photon energy of $E_L = 1.837$ eV (675 nm), photoemitted electrons can only be generated via a nonlinear photoemission process. The order of the photoemission process can be determined from the polarization-dependent PEEM yield by fitting a modified version of Malus´ law,

$$y(\alpha) = a + b(\cos{(\alpha + c)})^{2N}, \tag{S1}$$

where $N$ accounts for the nonlinearity of the photoemission process, $a$ is a residual signal offset, $b$ a yield scaling factor for the cosine function, and $c$ is an offset angle for the orientation angle $\alpha$. The fit results (solid lines in Fig. S5.1) yield $N = 3.2$ for both configurations, indicating that, on average, three absorbed photons are required for photoemission.

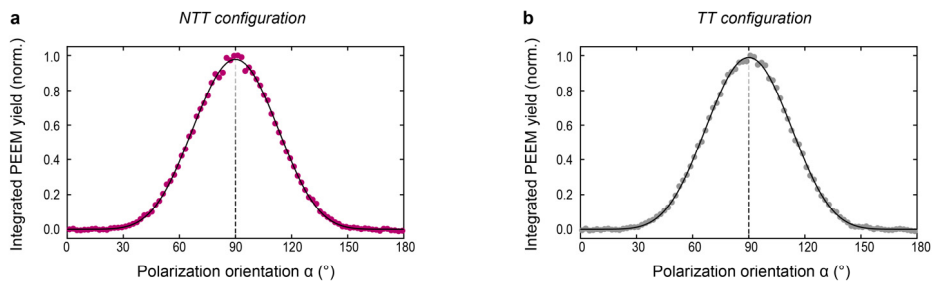

**Fig. S5.1: PEEM yield polarization orientation dependency. a,** Spatially integrated, background-subtracted, and normalized PEEM yield of the nontrivial configuration nanoslit chain (NTT, $v = -177$ meV and $w = -238$ meV) dependent on the polarization angle of the excitation light field with respect to the nanoslit chain orientation. Fit according to equation S1 (solid line) reveals an $N = 3$ order of the photoemission process. The dashed line indicates the PEEM yield maximum at $\alpha = 90°$, i.e., for a polarization along the short axis of nanoslit. **b,** Equivalent representation as in (a) for the trivial configuration (TT, $v = -238$ meV and $w = -177$ meV).